\PassOptionsToPackage{unicode}{hyperref}
\PassOptionsToPackage{hyphens}{url}
\PassOptionsToPackage{dvipsnames,svgnames,x11names}{xcolor}
\documentclass[
]{article}
\usepackage{amsmath,amssymb}
\usepackage{lmodern}
\usepackage{iftex}
\ifPDFTeX
  \usepackage[T1]{fontenc}
  \usepackage[utf8]{inputenc}
  \usepackage{textcomp} 
\else 
  \usepackage{unicode-math}
  \defaultfontfeatures{Scale=MatchLowercase}
  \defaultfontfeatures[\rmfamily]{Ligatures=TeX,Scale=1}
\fi
\IfFileExists{upquote.sty}{\usepackage{upquote}}{}
\IfFileExists{microtype.sty}{
  \usepackage[]{microtype}
  \UseMicrotypeSet[protrusion]{basicmath} 
}{}
\makeatletter
\@ifundefined{KOMAClassName}{
  \IfFileExists{parskip.sty}{%
    \usepackage{parskip}
  }{
    \setlength{\parindent}{0pt}
    \setlength{\parskip}{6pt plus 2pt minus 1pt}}
}{
  \KOMAoptions{parskip=half}}
\makeatother
\usepackage{xcolor}
\usepackage{color}
\usepackage{fancyvrb}

\DefineVerbatimEnvironment{Highlighting}{Verbatim}{commandchars=\\\{\}}
\newenvironment{Shaded}{}{}

\newcommand{\BuiltInTok}[1]{\textcolor[rgb]{0.00,0.50,0.00}{#1}}

\newcommand{\ControlFlowTok}[1]{\textcolor[rgb]{0.00,0.44,0.13}{\textbf{#1}}}

\newcommand{\DecValTok}[1]{\textcolor[rgb]{0.25,0.63,0.44}{#1}}

\newcommand{\FloatTok}[1]{\textcolor[rgb]{0.25,0.63,0.44}{#1}}

\newcommand{\ImportTok}[1]{\textcolor[rgb]{0.00,0.50,0.00}{\textbf{#1}}}

\newcommand{\KeywordTok}[1]{\textcolor[rgb]{0.00,0.44,0.13}{\textbf{#1}}}
\newcommand{\NormalTok}[1]{#1}
\newcommand{\OperatorTok}[1]{\textcolor[rgb]{0.40,0.40,0.40}{#1}}

\providecommand{\tightlist}{%
  \setlength{\itemsep}{0pt}\setlength{\parskip}{0pt}}
\NewDocumentCommand\citeproctext{}{}
\NewDocumentCommand\citeproc{mm}{%
  \begingroup\def\citeproctext{#2}\cite{#1}\endgroup}
\makeatletter
 \let\@cite@ofmt\@firstofone
 \def\@biblabel#1{}
 \def\@cite#1#2{{#1\if@tempswa , #2\fi}}
\makeatother
\newlength{\cslhangindent}
\newlength{\csllabelwidth}
\newenvironment{CSLReferences}[2] 
 {\begin{list}{}{%
  \setlength{\itemindent}{0pt}
  \setlength{\leftmargin}{0pt}
  \setlength{\parsep}{0pt}
  \ifodd #1
   \setlength{\leftmargin}{\cslhangindent}
   \setlength{\itemindent}{-1\cslhangindent}
  \fi
  \setlength{\itemsep}{#2\baselineskip}}}
 {\end{list}}
\usepackage{calc}

\ifLuaTeX
\usepackage[bidi=basic]{babel}
\else
\usepackage[bidi=default]{babel}
\fi
\babelprovide[main,import]{american}

\def\languageshorthands#1{}
\ifLuaTeX
  \usepackage{selnolig}  
\fi
\IfFileExists{bookmark.sty}{\usepackage{bookmark}}{\usepackage{hyperref}}
\IfFileExists{xurl.sty}{\usepackage{xurl}}{} 
\hypersetup{
  pdftitle={lintsampler: Easy random sampling via linear interpolation},
  pdfauthor={Aneesh P. Naik, Michael S. Petersen},
  pdflang={en-US},
  colorlinks=true,
  linkcolor={Maroon},
  filecolor={Maroon},
  citecolor={Blue},
  urlcolor={Blue},
  pdfcreator={LaTeX via pandoc}}

\title{lintsampler: Easy random sampling via linear interpolation}

\definecolor{c53baa1}{RGB}{83,186,161}
\definecolor{c202826}{RGB}{32,40,38}

\usepackage[affil-it]{authblk}
\usepackage{orcidlink}
\author[1%
  \ensuremath\mathparagraph]{Aneesh P. Naik%
    \,\orcidlink{0000-0001-6841-1496}\,%
    }
\author[1%
  ]{Michael S. Petersen%
    \,\orcidlink{0000-0003-1517-3935}\,%
    }

\affil[1]{Institute for Astronomy, University of Edinburgh, UK%
  }
\affil[$\mathparagraph$]{Corresponding author: %
}
\date{14 June 2024}

\begin{document}
\maketitle

\section{Summary}\label{summary}

\texttt{lintsampler} provides a Python implementation of a technique we
term `linear interpolant sampling': an algorithm to efficiently draw
pseudo-random samples from an arbitrary probability density function
(PDF). First, the PDF is evaluated on a grid-like structure. Then, it is
assumed that the PDF can be approximated between grid vertices by the
(multidimensional) linear interpolant. With this assumption, random
samples can be efficiently drawn via inverse transform sampling
(\citeproc{ref-devroye.book}{Devroye, 1986}).

\texttt{lintsampler} is primarily written with \texttt{numpy}
(\citeproc{ref-numpy.paper}{Harris et al., 2020}), drawing some
additional functionality from \texttt{scipy}
(\citeproc{ref-scipy.paper}{Virtanen et al., 2020}). Under the most
basic usage of \texttt{lintsampler}, the user provides a Python function
defining the target PDF and some parameters describing a grid-like
structure to the \texttt{LintSampler} class, and is then able to draw
samples via the \texttt{sample} method. Additionally, there is
functionality for the user to set the random seed, employ quasi-Monte
Carlo sampling, or sample within a premade grid (\texttt{DensityGrid})
or tree (\texttt{DensityTree}) structure.

\section{Statement of need}\label{statement-of-need}

Below is a (non-exhaustive) list of `use cases', i.e., situations where
a user might find \texttt{lintsampler} (and/or the the linear
interpolant sampling algorithm underpinning it) to be preferable over
random sampling techniques such as importance sampling, rejection
sampling or Markov chain Monte Carlo (MCMC). MCMC in particular is a
powerful class of methods with many excellent Python implementations
(\citeproc{ref-sgmcmcjax.paper}{Coullon \& Nemeth, 2022};
\citeproc{ref-emcee.paper}{Foreman-Mackey et al., 2019};
\citeproc{ref-pxmcmc.paper}{Marignier, 2023};
\citeproc{ref-pymc.paper}{Patil et al., 2010}). In certain use cases as
described below, \texttt{lintsampler} can offer more convenient and/or
more efficient sampling compared with these various techniques.

We'll assume that the target PDF the user wishes to sample from does not
have its own exact sampling algorithm (such as the Box-Muller transform
for a Gaussian PDF). The power of \texttt{lintsampler} lies in its
applicability to arbitrary PDFs for which tailor-made sampling
algorithms are not available.

\subsection{Use Cases}\label{use-cases}

\subsubsection{1. Expensive PDF}\label{expensive-pdf}

If the PDF being sampled from has high computatational overhead to
evaluate (referred to as computationally `expensive') and a large number
of samples is desired, then \texttt{lintsampler} might be the most
cost-effective option. This is because \texttt{lintsampler} does not
evaluate the PDF for each sample (as would be the case for other random
sampling techniques), but on the nodes of the user-chosen grid.
Particularly in a low-dimensional setting where the grid does not have
too many nodes, this can mean far fewer PDF evaluations. This point is
demonstrated in the
\href{https://lintsampler.readthedocs.io/en/latest/example_notebooks/1_gmm.html}{first
example notebook} in the \texttt{lintsampler} docs.\footnote{Similarly,
  there might be situations where the user is not so concerned about
  strict statistical representativeness but wants to generate a huge
  number of samples from a target PDF with the least possible
  computational cost (such as e.g., generating realistic point cloud
  distributions in video game graphics). They can use
  \texttt{lintsampler} with a coarse grid (so minimal PDF evaluations),
  then \texttt{sample()} to their heart's content.}

\subsubsection{2. Multimodal PDF}\label{multimodal-pdf}

If the target PDF has a highly complex structure with multiple,
well-separated modes, then \texttt{lintsampler} might be the
\emph{easiest} option (in terms of user configuration). In such
scenarios, MCMC might struggle unless the walkers are carefully
preconfigured to start near the modes. Similarly, rejection sampling or
importance sampling would be highly suboptimal unless the proposal
distribution is carefully chosen to match the structure of the target
PDF. With \texttt{lintsampler}, the user needs only to ensure that the
resolution of their chosen grid is sufficient to resolve the PDF
structure, and so the setup remains straightforward. This is
demonstrated in the
\href{https://lintsampler.readthedocs.io/en/latest/example_notebooks/2_doughnuts.html}{second
example notebook} in the \texttt{lintsampler} docs.\footnote{It is worth
  noting that in these kinds of complex, multimodal problems, a single
  fixed grid might not be the most cost-effective sampling domain. For
  this reason, \texttt{lintsampler} also provides simple functionality
  for sampling over multiple disconnected grids.}

\subsubsection{3. PDF with large dynamic
range}\label{pdf-with-large-dynamic-range}

If the target PDF has a very large dynamic range, then the
\texttt{DensityTree} object provided by \texttt{lintsampler} might be an
effective solution. Here, the PDF is evaluated not on a fixed grid, but
on the leaves of a tree. The tree is refined such that regions of
concentrated probability are more finely resolved, based on accuracy
criteria. Such an example is shown in the
\href{https://lintsampler.readthedocs.io/en/latest/example_notebooks/3_dark_matter.html}{third
example notebook} in the \texttt{lintsampler} docs.

\subsubsection{4. Noise needs to be
minimised}\label{noise-needs-to-be-minimised}

In Quasi-Monte Carlo (QMC) sampling, one purposefully generates more
`balanced' (and thus less random) draws from a target PDF, so that
sampling noise decreases faster than \(\mathcal{O}(N^{-1/2})\).
\texttt{lintsampler} allows easy QMC sampling with arbitrary PDFs. We
are not aware of such capabilities with any other package. We give an
example of using \texttt{lintsampler} for QMC in the
\href{https://lintsampler.readthedocs.io/en/latest/example_notebooks/4_qmc.html}{fourth
example notebook} in the \texttt{lintsampler} docs.

\subsection{`Real World' Example}\label{real-world-example}

Any one of the four use cases above would serve by itself as a
sufficient case for choosing \texttt{lintsampler}, but here we give an
example of a real-world scenario that combines all of the use cases. It
is drawn from our own primary research interests in computational
astrophysics.

Much of computational astrophysics consists of large-scale high
performance computational simulations of gravitating systems. For
example, simulations of planets evolving and interacting within a solar
system, simulations of stars interacting within a galaxy, or vast
cosmological simulations in which a whole universe is grown \textbf{in
silico}.

There exists a myriad of codes used to run these simulations, each using
different algorithms to solve the governing equations. One class of
simulation code that has gained much attention in recent years is the
class of code employing basis function expansions
(\citeproc{ref-exp.paper}{Petersen et al., 2022};
\citeproc{ref-agama.paper}{Vasiliev, 2019}). In these codes, the matter
density at any point in space is represented by a sum over basis
functions (not unlike a Fourier series), with the coefficients in the
sum changing over space and time. As such, the matter comprising the
system is represented everywhere as a smooth, continous fluid, but for
many applications and/or downstream analyses of the simulated system,
one needs to instead represent the system as a set of discrete
particles. These particles can be obtained by drawing samples from the
continuous density field.

This is a scenario that satisfies all four of the use cases list above.
To explain further:

\begin{itemize}
\tightlist
\item
  The PDF we are sampling from (i.e., the basis expansion representation
  of the matter density field) can be expensive to evaluate if a large
  number of terms are included in the sum.
\item
  The PDF can be highly multimodal when the system we are simulating
  comprises many distinct gravitating substructures, such as stellar
  clusters.
\item
  The PDF can have a large dynamic range. Astrophysical structures such
  as galaxies and dark matter `haloes' often have power-law density
  profiles, such as the Navarro-Frenk-White profile
  (\citeproc{ref-nfw.paper}{Navarro et al., 1997}). Further complicating
  the issue is that a typical dark matter halo will host several
  `subhaloes', which in turn might host `subsubhaloes', and so on. In
  short, a range of spatial scales needs to be resolved.
\item
  If the particle set being sampled is to be used for further
  simulation, it can be helpful to draw as `noiseless' a sample as
  possible for reasons of numerical stability.
\end{itemize}

For these reasons, this kind of astrophysical simulation code provides
an excellent example of a `real world' application for
\texttt{lintsampler}. Here, one would cover the simulation domain with a
\texttt{DensityTree} instance (or several instances, one for each
primary structure), call the \texttt{refine} method to better resolve
the high-density regions, then feed the tree to a \texttt{LintSampler}
instance and call \texttt{sample} to generate particles. The
\texttt{qmc} flag can be passed to the sampler in order to employ
Quasi-Monte Carlo sampling.

\subsection{Caveats}\label{caveats}

In all use cases listed above, it is assumed that the dimension of the
problem is not too high. \texttt{lintsampler} works by evaluating a
given PDF on the nodes of a grid (or grid-like structure, such as a
tree), so the number of evaluations (and memory occupancy) grows
exponentially with the number of dimensions. As a consequence, many of
the efficiency arguments given for \texttt{lintsampler} below don't
apply to higher dimensional problems. We probably wouldn't use
\texttt{lintsampler} in more than 6 dimensions, but there is no hard
limit here: the question of how many dimensions is too many will depend
on the problem at hand.

\section{Usage}\label{usage}

\texttt{lintsampler} is designed with an interface that makes sampling
from an input PDF straightforward. For example, if you have PDF with
multiple separated peaks:

\begin{Shaded}
\begin{Highlighting}[]
\ImportTok{import}\NormalTok{ numpy }\ImportTok{as}\NormalTok{ np}
\ImportTok{from}\NormalTok{ scipy.stats }\ImportTok{import}\NormalTok{ norm}

\KeywordTok{def}\NormalTok{ gmm\_pdf(x):}
\NormalTok{    mu }\OperatorTok{=}\NormalTok{ np.array([}\OperatorTok{{-}}\FloatTok{3.0}\NormalTok{, }\FloatTok{0.5}\NormalTok{, }\FloatTok{2.5}\NormalTok{])}
\NormalTok{    sig }\OperatorTok{=}\NormalTok{ np.array([}\FloatTok{1.0}\NormalTok{, }\FloatTok{0.25}\NormalTok{, }\FloatTok{0.75}\NormalTok{])}
\NormalTok{    w }\OperatorTok{=}\NormalTok{ np.array([}\FloatTok{0.4}\NormalTok{, }\FloatTok{0.25}\NormalTok{, }\FloatTok{0.35}\NormalTok{])}
    \ControlFlowTok{return}\NormalTok{ np.}\BuiltInTok{sum}\NormalTok{([w[i] }\OperatorTok{*}\NormalTok{ norm.pdf(x, mu[i], sig[i]) }\ControlFlowTok{for}\NormalTok{ i }\KeywordTok{in} \BuiltInTok{range}\NormalTok{(}\DecValTok{3}\NormalTok{)], axis}\OperatorTok{=}\DecValTok{0}\NormalTok{)}
\end{Highlighting}
\end{Shaded}

\texttt{lintsampler} can efficiently draw samples from it on some
defined interval:

\begin{Shaded}
\begin{Highlighting}[]
\ImportTok{from}\NormalTok{ lintsampler }\ImportTok{import}\NormalTok{ LintSampler}

\NormalTok{grid }\OperatorTok{=}\NormalTok{ np.linspace(}\OperatorTok{{-}}\DecValTok{7}\NormalTok{,}\DecValTok{7}\NormalTok{,}\DecValTok{100}\NormalTok{)}
\NormalTok{samples }\OperatorTok{=}\NormalTok{ LintSampler(grid,pdf}\OperatorTok{=}\NormalTok{gmm\_pdf).sample(N}\OperatorTok{=}\DecValTok{10000}\NormalTok{)}
\end{Highlighting}
\end{Shaded}

\texttt{samples} is then an array of 10000 samples drawn from the PDF.
Apart from defining the PDF, \texttt{lintsampler} enables creating
discrete samples from a continuous PDF in a small handful of lines.

\section{Features}\label{features}

Although \texttt{lintsampler} is written in pure Python, making the code
highly readable, the methods make extensive use of \texttt{numpy}
functionality to provide rapid sampling. After the structure spanning
the domain has been constructed, sampling proceeds with computational
effort scaling linearly with number of sample points.

We provide two methods to define the domain, both optimised with
\texttt{numpy} functionality for efficient construction. The
\texttt{DensityGrid} class takes highly flexible inputs for defining a
grid. In particular, the grid need not be evenly spaced (or even
continuous) in any dimension; the user can preferentially place grid
elements near high-density regions. The \texttt{DensityTree} class takes
error tolerance parameters and constructs an adaptive structure to
achieve the specified tolerance. We also provide a base class
(\texttt{DensityStructure}) such that the user could extend the methods
for spanning the domain.

Documentation for \texttt{lintsampler}, including example notebooks
demonstrating a range of problems, is available via a
\href{https://lintsampler.readthedocs.io}{readthedocs page}. The
documentation also has an extensive explanation of the interfaces,
including optimisation parameters for increasing the efficiency in
sampling.

\section{Acknowledgements}\label{acknowledgements}

We would like to thank Sergey Koposov for useful discussions. APN
acknowledges funding support from an Early Career Fellowship from the
Leverhulme Trust. MSP acknowledges funding support from a UKRI Stephen
Hawking Fellowship.

\section*{References}\label{references}
\addcontentsline{toc}{section}{References}

\phantomsection\label{refs}

\end{document}